\documentclass[twoside,twocolumn,9pt]{article}
\usepackage{extsizes}
\usepackage[super,sort&compress,comma]{natbib} 
\usepackage[version=3]{mhchem}
\usepackage[left=1.5cm, right=1.5cm, top=1.785cm, bottom=2.0cm]{geometry}
\usepackage{balance}
\usepackage{mathptmx}
\usepackage{sectsty}
\usepackage{graphicx} 
\usepackage{lastpage}
\usepackage[format=plain,justification=justified,singlelinecheck=false,font={stretch=1.125,small,sf},labelfont=bf,labelsep=space]{caption}
\usepackage{float}
\usepackage{fancyhdr}
\usepackage{fnpos}
\usepackage[english]{babel}
\addto{\captionsenglish}{%
	\renewcommand{\refname}{Notes and references}
}
\usepackage{array}
\usepackage{droidsans}
\usepackage{charter}
\usepackage[T1]{fontenc}
\usepackage[usenames,dvipsnames]{xcolor}
\usepackage{setspace}
\usepackage[compact]{titlesec}
\usepackage{hyperref}

\usepackage{lineno,subfig,siunitx} 

\usepackage{epstopdf}

\definecolor{cream}{RGB}{222,217,201}

\begin{document}
	
	\pagestyle{fancy}
	\thispagestyle{plain}
	\fancypagestyle{plain}{
		\renewcommand{\headrulewidth}{0pt}
	}
	
	\makeFNbottom
	\makeatletter
	\renewcommand\LARGE{\@setfontsize\LARGE{15pt}{17}}
	\renewcommand\Large{\@setfontsize\Large{12pt}{14}}
	\renewcommand\large{\@setfontsize\large{10pt}{12}}
	\renewcommand\footnotesize{\@setfontsize\footnotesize{7pt}{10}}
	\makeatother
	
	\renewcommand{\thefootnote}{\fnsymbol{footnote}}
	\renewcommand\footnoterule{\vspace*{1pt}%
		\color{cream}\hrule width 3.5in height 0.4pt \color{black}\vspace*{5pt}} 
	\setcounter{secnumdepth}{5}
	
	\makeatletter 
	\renewcommand\@biblabel[1]{#1}            
	\renewcommand\@makefntext[1]%
	{\noindent\makebox[0pt][r]{\@thefnmark\,}#1}
	\makeatother 
	\renewcommand{\figurename}{\small{Fig.}~}
	\sectionfont{\sffamily\Large}
	\subsectionfont{\normalsize}
	\subsubsectionfont{\bf}
	\setstretch{1.125} 
	\setlength{\skip\footins}{0.8cm}
	\setlength{\footnotesep}{0.25cm}
	\setlength{\jot}{10pt}
	\titlespacing*{\section}{0pt}{4pt}{4pt}
	\titlespacing*{\subsection}{0pt}{15pt}{1pt}
	
	\fancyfoot{}
	\fancyfoot[LO,RE]{\vspace{-7.1pt}}
	\fancyfoot[RO]{\footnotesize{\sffamily{\thepage}}}
	\fancyfoot[LE]{\footnotesize{\sffamily{\thepage}}}
	\fancyhead{}
	\renewcommand{\headrulewidth}{0pt} 
	\renewcommand{\footrulewidth}{0pt}
	\setlength{\arrayrulewidth}{1pt}
	\setlength{\columnsep}{6.5mm}
	\setlength\bibsep{1pt}
	
	\makeatletter 
	\newlength{\figrulesep} 
	\setlength{\figrulesep}{0.5\textfloatsep} 
	
	\newcommand{\topfigrule}{\vspace*{-1pt}%
		\noindent{\color{cream}\rule[-\figrulesep]{\columnwidth}{1.5pt}} }
	
	\newcommand{\botfigrule}{\vspace*{-2pt}%
		\noindent{\color{cream}\rule[\figrulesep]{\columnwidth}{1.5pt}} }
	
	\newcommand{\dblfigrule}{\vspace*{-1pt}%
		\noindent{\color{cream}\rule[-\figrulesep]{\textwidth}{1.5pt}} }
	
	\makeatother
	
	\twocolumn[
	\begin{@twocolumnfalse}
		\vspace{1em}
		\sffamily
		\begin{tabular}{m{1.5cm} p{16.5cm} }

			 & \noindent\LARGE{\textbf{Thermal and electrical cross-plane conductivity at the nanoscale in poly(3,4-ethylenedioxythiophene):trifluoromethanesulfonate thin films$^\dag$
				}} \\
			\vspace{0.3cm} & \vspace{0.3cm} \\
			
			 & \noindent\large{Kirill Kondratenko,$^{\ast}$\textit{$^{a}$} David Gu\'erin,\textit{$^{a}$} Xavier Wallart,\textit{$^{a}$} St\'ephane Lenfant\textit{$^{a}$} and Dominique Vuillaume$^{\ast}$\textit{$^{a}$}
			} \\

			& \noindent\normalsize{
				Cross-plane electrical and thermal transport in thin films of a conducting polymer (poly(3,4-ethylenedioxythiophene), PEDOT) stabilized with trifluoromethanesulfonate (OTf) is investigated in this study. We explore their electrical properties by conductive atomic force microscopy (C-AFM), which reveals the presence of highly conductive nano-domains. 
				Thermal conductivity in cross-plane direction is measured with Null-Point scanning thermal microscopy (NP-SThM): PEDOT:OTf indeed demonstrates non-negligible electronic contribution to the thermal transport. We further investigate the correlation between electrical and thermal conductivity by applying post-treatment: chemical reduction (de-doping) for the purpose of lowering charge carrier concentration and hence, electrical conductivity and acid treatment (over-doping) to increase the latter.
				From our measurements, we find a vibrational thermal conductivity of 0.34±0.04 \si{\W\per\m\per\K}. From the linear dependence or the electronic contribution of thermal conductivity vs. the electronic conductivity (Widemann-Franz law), we infer a Lorenz number ~6 times larger than the classical Sommerfeld value as also observed in many organic materials for in-plane thermal transport. Applying the recently proposed molecular Widemann-Franz law, we deduced a reorganization energy of \SI[separate-uncertainty=true,multi-part-units=single]{0.53\pm0.06}{\eV}.
			} \\
			
		\end{tabular}
		
	\end{@twocolumnfalse} \vspace{0.6cm}
	
	]
	
	\renewcommand*\rmdefault{bch}\normalfont\upshape
	\rmfamily
	\section*{}
	\vspace{-1cm}

	
	\footnotetext{\textit{$^{a}$~Institut d'Electronique Micro\'electronique et Nanotechnologie (IEMN), CNRS, Villeneuve d'Ascq, France. }}
	\footnotetext{\textit{$^{\ast}$~E-mail: kirill.kondratenko@iemn.fr, dominique.vuillaume@iemn.fr }}
	
	\footnotetext{\dag~Electronic Supplementary Information (ESI) available: details on polymer film deposition, thermal and electrical characterization, topography images in tapping mode, oxidation degree calculation from XPS data and NP SThM data. See DOI: 00.0000/00000000.}
	
%
	
	


\section{Introduction}

Conducting polymers are recognized as promising candidates for varying applications: flexible and printable circuits \cite{Viola2020}, transistors \cite{Chang2004}, light-emitting diodes \cite{Devices2006}, solar cells \cite{Helgesen2010}. As potential materials for low cost thermoelectric (TE) devices, conducting polymers occupy special place: historically, significant efforts were directed into this field of research due to intrinsically low thermal conductivity of organic materials \cite{Wang2012b}, \cite{Toshima2017}, \cite{Bubnova2011}
. The performance of a TE material is typically evaluated by the value of its dimensionless figure-of-merit, $ZT$:
\begin{equation}
	\centering
	ZT=\frac{S^2 \sigma}{\kappa_v + \kappa_e} T, \label{eq:ZT}
\end{equation}
where $S$ is the Seebeck coefficient, $\sigma$ is the electrical conductivity, $\kappa_v$ is the vibrational thermal conductivity, $\kappa_e$ is the electronic thermal conductivity and $T$ is the temperature. 
In addition to high electrical conductivity, a suitable TE material is required to have low thermal conductivity in order to achieve high ZT values \cite{Shi2017a}, which makes the understanding of electronic contribution to thermal transport a subject of paramount importance to this field of studies.

Recently, thin films of poly(3,4-ethylenedioxythiophene) (PEDOT) stabilized with trifluoromethanesulfonate (triflate, OTf) counter-ions have demonstrated exceptionally high electrical conductivity exceeding 3000 \si{\siemens\per\centi\meter} \cite{Gueye2016}, which makes it an excellent candidate for potential TE applications. This material presents a significant degree of control over its TE properties (optimization of ZT up to 0.25) through dopant content modulation, as demonstrated by Bubnova \textit{et al} \cite{Bubnova2011}. Kim \textit{et al} have achieved ZT of 0.42 at room temperature for films of PEDOT stabilized with polystyrene sulfonate (PSS) \cite{Kim2013b} by a post-deposition treatment with ethylene glycole.

Considerable attention was brought to investigation of thermal properties of conductive polymers due to the aforementioned interest in thermoelectric applications \cite{Fan2019}. Moreover, the development of active circuit elements incorporating organic substances has spurred interest in the investigations of thermal conductivity due to heat dissipation limitations of integrated circuits. 
In relation to thermoelectric materials, the electronic contribution to thermal conductivity, which arises from the increase of electrical conductivity is generally considered undesirable. Therefore, additional effort should be put into understanding the mechanism which connects electronic and thermal transport in conducting properties.

In addition to doping \cite{Kondratenko2019a}, control over crystallinity presents itself as a powerful strategy of improving electrical conductivity by means of increased charge carrier mobility in conducting polymers, as demonstrated by Cho \textit{et al.} \cite{Cho2014} for PEDOT nanowires. Gueye \textit{et al.} \cite{Gueye2016} 
have concluded from X-ray difractograms that samples of PEDOT stabilized with small counter-ions (triflate and sulfonate) had improved crystallinity which resulted in a more densely packed polymer structure and enhanced charge transport demonstrating semi-metallic behavior \cite{Rudd2018} as compared to commonly known PEDOT:PSS.
%


In this work we explore electrical and thermal properties of thin films of PEDOT:OTf in cross-plane direction using scanning probe microscopy (SPM) techniques. Electrical conductivity of polymer material is investigated with conductive atomic force microscopy (C-AFM), where electrical current through the sample between the conductive AFM tip and the underlying substrate is measured. Electrical current imaging reveals presence of highly conductive domains in PEDOT:OTf films corresponding to the crystallites with increased interchain order which is responsible for the favorable transporting properties.

We used scanning thermal microscopy (SThM) for the characterization of local thermal properties of PEDOT:OTf. This technique is frequently applied to a large variety of samples, such as organic thin films \cite{Gueye2021}, \cite{Trefon-Radziejewska2017}, \cite{Zhang2019a}, nanomaterials \cite{Chen2020}, \cite{Tortello2019}, \cite{Konemann2019} and single molecules \cite{Meier2014}, \cite{Mosso2019}. In the SThM setup, the thin film resistor in the tip of the SThM cantilever is essentially utilized as a thermometer in order to probe temperature with nanometric lateral resolution (in the so-called passive mode) or to induce heat flow from the tip into the sample by supplying power to the aforementioned resistor (in the active mode), which can be used to find the thermal conductance of materials under study \cite{Zhang2019}, \cite{Kondratenko2021a}. 

Here, we study the dependence of thermal conductivity versus the electrical transporting properties of PEDOT samples. We are able to tune their conductivity by chemically reducing the polymeric film which results in a decrease of conductivity. On the other hand, we employ secondary doping with sulfuric acid to increase the conductivity of as-deposited PEDOT:OTf. Combination of these approaches allows us to vary the electrical conductivity in a span of 4 orders of magnitude. We extract the vibrational thermal conductivity $\kappa_v$ of PEDOT:OTF from the plot of thermal conductivity as a function of electrical conductivity. Furthermore, we are able to estimate the Lorenz number value for this material as well as to provide an estimate of the energy of molecular reorganization for a PEDOT chain according to the recently developed molecular Wiedemann-Franz (MWF) law \cite{Craven2020}. 

\section{Experimental}

\subsection{Sample preparation}
PEDOT:OTf thin films were prepared according to the procedure described elsewhere \cite{Gueye2016} on clean substrates (Si wafer covered with sputtered Ti/Au 20/200 nm). We have employed N,N-dimethylformamide (DMF) and N-methyl-2-pyrrolidone (NMP) as co-solvents in our precursor solution formulation in order to increase electrical conductivity as described by Yvenou \textit{et al} \cite{Yvenou2020}.  See Section 1 of Supporting Information (SI) for the complete preparation procedure.

\subsection{Spectroscopic characterization}

XPS measurements were performed to analyze the chemical composition of the thin films and to estimate the doping degree of PEDOT. We used a Physical Electronics 5600 spectrometer fitted in an UHV chamber with a residual pressure of \SI{3e-10}{\milli\bar}. Confocal micro-Raman spectrometer (Horiba-Jobin Yvon, LabRam®HR) was utilized for Raman spectra acquisition. A diode pumped solid state laser (Cobalt Blues) with a wavelength of 473 nm (maximum output power of 25 mW) was used as an excitation source. Raman signal was collected in back-scattering configuration with a x100 objective lens and dispersed by a 1800 lines/mm grating before being directed to a CCD detector.

\subsection{Electrical characterization}

C-AFM measurements on as-prepared and doped with sulfuric acid samples were performed in air at T=\SI{22.5}{\degreeCelsius} and relative humidity of 35-40\% on a Bruker Dimension Icon AFM. Measurements were performed in contact mode with a PtIr coated tip (Bruker SCM PIC V2, spring constant 0.1 \si{\newton\per\meter}). Measurements on chemically reduced materials were performed in ultra-high vacuum (UHV) conditions (\SI{e-10}{} to \SI{e-9}{\milli\bar}) in order to preserve the oxidation degree of the sample prior to the subsequent NP-SThM measurements. For these measurements, VT-SPM microscope (Scienta Omicron) with RMN-12PT400B probe (solid platinum, spring constant 0.3 \si{\newton\per\meter}) was used. Loading force was maintained between 10 and 20 nN during the measurements for both types of characterization.  Details on calculation of cross-plane electrical conductivity are provided in SI (Section 2).

\begin{figure*}[h!]
	\centering
	\includegraphics[width=0.99\textwidth]{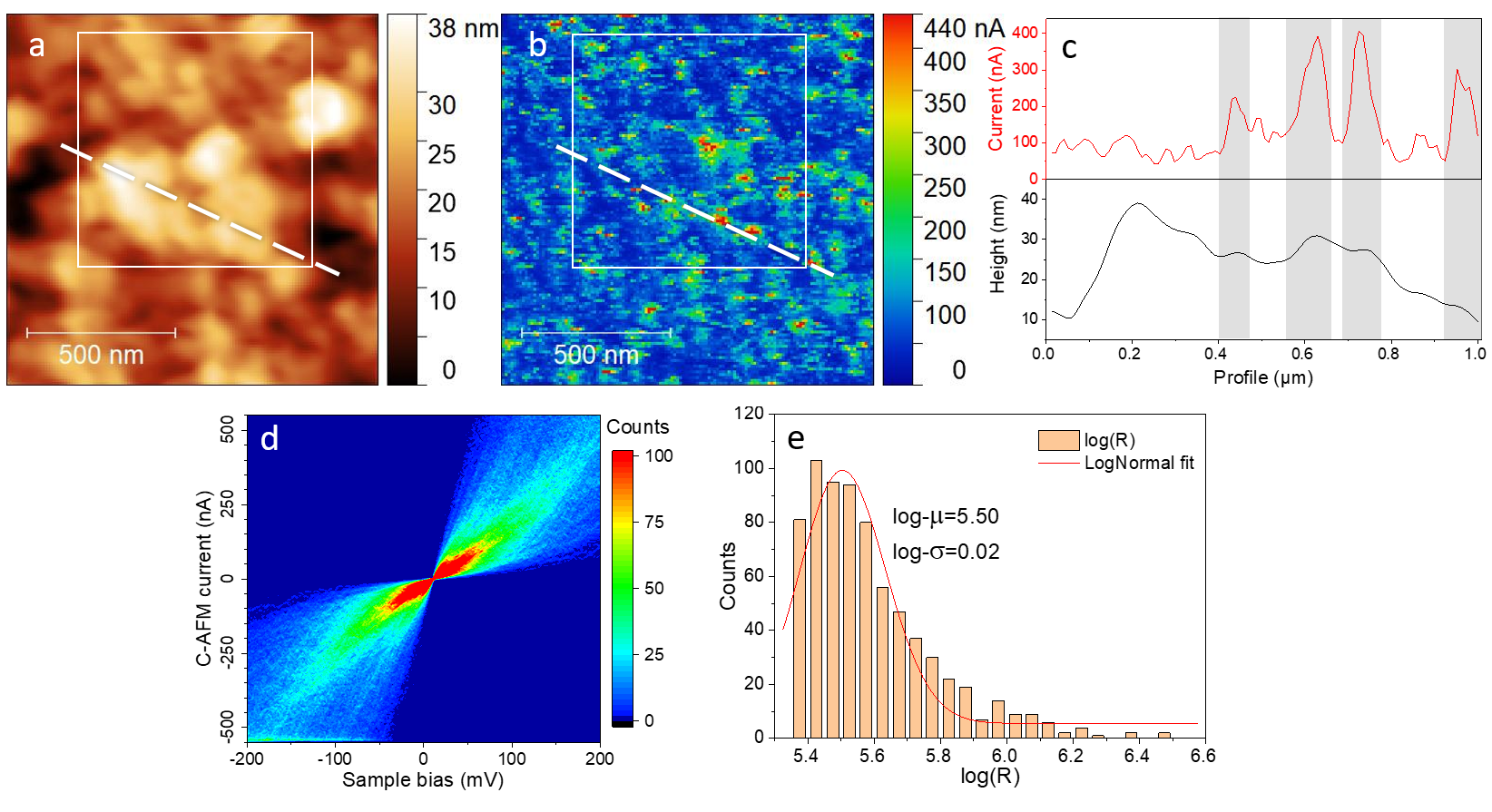}
	\caption[]{C-AFM measurements of PEDOT:OTf sample: (a) Topography of a 1.25$\times$1.25 \si{\micro\meter} area of polymer film in contact mode (12 nN constant force). (b) Electrical current image at 50 mV sample bias. (c) Profiles extracted from images (a) and (b) (marked with a solid line): current on the top, topography on the bottom. (d) 2D histogram of 720 I(V) measurements in the square area highlighted in the images (a) and (b) with a dashed line. (e) Log(Resistance) map of the highlighted area (color scale adjusted for clarity). 
	}
	\label{fig:CAFM_map}
\end{figure*}

\subsection{Thermal conductivity measurements}

Nanoscale thermal characterization of PEDOT-based materials was carried out in air on a Bruker Dimension Icon AFM with Anasys SThM module.  We used Kelvin NanoTechnology (KNT) SThM probes with Pd thin film resistor in the probe tip as the heating element (VITA-HE-GLA-1, spring constant 0.5 \si{\newton\per\meter}). This probe has increased contact area and thus a larger probed volume for SThM (see Section 3 of SI) when compared to a classic contact mode AFM probe (SThM probe tip radius  $\sim$\SI{100}{\nano\meter}) which is necessary to attain a measurable heat flow from the tip into the sample \cite{Zhang2019}.

Cross-plane thermal conductivity was estimated with the Null Point (NP) technique. This differential method allows to obtain quantitative information on thermal conductivity by subtracting the parasitic heat loss from the probe to the environment \cite{Kim2011b}. The measurements are performed by approaching the sample surface with a heated SThM probe until the physical contact is made. The tip temperature is recorded during the cantilever travel: before contact, it is primarily influenced by losses through the air and the body of SThM probe; at the moment of contact the heat flow through the tip-sample interface arises and is reflected in the total heat dissipation of the probe. This rapid temperature change due to the aforementioned contact (T$_{NC}$-T$_C$) is recorded for a set of tip temperatures alongside with the contact temperature T$_C$, which is then used in the following equation (\ref{eq:np_sthm}) in order to extract the thermal conductivity $\kappa$:
\begin{equation}
	T_C-T_{amb} = \left[ \alpha \frac{1}{\kappa} + \beta\right] (T_{NC}-T_C) \label{eq:np_sthm}
\end{equation}
where T$_C$ is the probe tip temperature at contact, i.e. when the tip is in thermal equilibrium with the sample surface, T$_{NC}$ is the tip temperature just before contact with the sample surface,  T$_{amb}$ is the ambient temperature, the calibration coefficients $\alpha $ and $\beta$ are related to thermal contact area, i.e. tip and sample geometry as well as tip-sample thermal conductance and the parasitic heat flow \cite{Kim2011b}. See Section 3 of SI for the details on calibration procedure.

%

\section{Results}

\subsection{Nanoscale electrical transport in pristine PEDOT:OTf} \label{section:Nanoscale_props} 

\begin{figure*}[h!]
	\centering
	
	\includegraphics[width=0.99\textwidth]{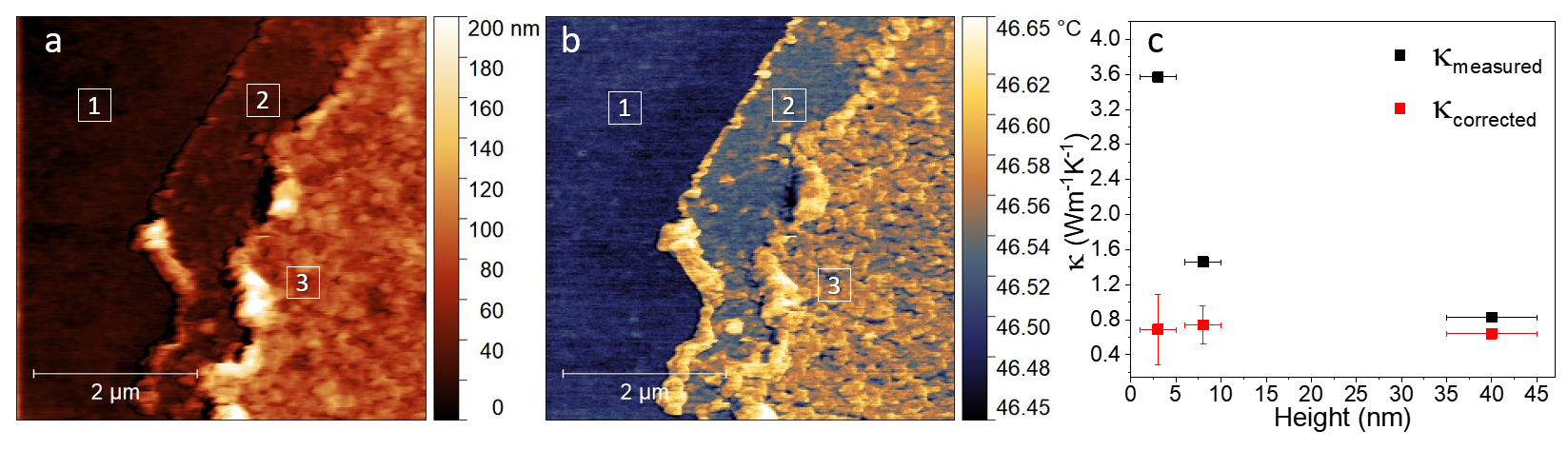}
	
	\caption[]{SThM characterization of pristine PEDOT:OTf film: (a) Topography in the contact mode of a 5$\times$5 \si{\micro\meter} area (constant force of ~20nN). (b) Temperature image recorded simultaneously with the topographical image at 0.7 V Wheatstone bridge value. (c) Plot of thermal conductivity as a function of PEDOT:OTf layer thickness measured in the areas highlighted by solid rectangles in the images (a) and (b). Black symbols correspond to the as-measured value, red symbols correspond to the values adjusted for the substrate contribution.}
	\label{fig:SThM_props}
\end{figure*}

As-deposited, PEDOT:OTf film presents granular structure with grain size in order of several tenths of nanometers which translates to relatively elevated roughness for a polymeric material (root mean square (RMS) roughness of about 4 nm, see Figure S2 in SI). This morphology is the direct consequence of the deposition method: in contrast to  commonly employed PEDOT:PSS, where spin coating of polymer solution results in a more smooth surface (RMS roughness in order of 1 nm \cite{Yemata2020a}), PEDOT:OTf is polymerized \textit{in situ} on the surface of the substrate from many nucleation sites in the precursor solution. Typical thickness for these films was about 40 to 45 nanometers (measured with AFM, see Section 5 of SI), this deposition method does not allow to obtain uniform films thicker than $\sim$\SI{50}{\nano\meter}. 

Electrical transport in PEDOT:OTf films was investigated with C-AFM. Figures~\ref{fig:CAFM_map} (a) and (b) demonstrate topography in contact mode (see Figure S4 of SI for higher resolution tapping mode AFM image) and current image recorded at 50 mV sample bias for a 1.25$\times$1.25 \si{\micro\meter} area of a PEDOT:OTf sample prepared with DMF co-solvent. It is of interest to point out that the amplitude from current image does not always show direct correlation to the film thickness: Figure~\ref{fig:CAFM_map} (c) demonstrates the comparison of topology and current intensity for the cross-section highlighted in images (a) and (b) of the same Figure. Current spikes over 200 nA are highlighted in gray: we can see in the topology profile (bottom part of the Figure~\ref{fig:CAFM_map} (c)) that these highly conductive areas indeed correspond to granules of PEDOT:OTf material and not to pits in the polymer film (similar to results presented by Osaka \textit{et al} \cite{Osaka2013}). We may also note that this behavior is not related to the variation of contact between C-AFM tip and the sample, as some other grains of PEDOT:OTf film demonstrate current amplitude comparable to the surrounding material. High Resolution Transmission Electron Microscopy (HRTEM) studies by Gueye \textit{et al.} \cite{Gueye2016} have demonstrated the presence of highly oriented crystallites in the films of PEDOT:OTf. 
The size of these crystallites (in the order of 10 nm) allows us to suggest that highly conductive grains of PEDOT:OTf film have increased content of these crystallites.

%
%

We have performed measurements of the electrical conductivity in cross-plane direction of the PEDOT:OTf film in the area highlighted in the images (a) and (b) of the Figure~\ref{fig:CAFM_map} by the white dashed line. 720 I(V) ramps from -200 to 200 mV were recorded in a 12$\times$12 grid with a 60 nm pitch, 5 ramps were recorded at each point. Figure~\ref{fig:CAFM_map} (d) shows a 2D histogram of current-voltage curves: PEDOT:OTf demonstrates ohmic conduction which is expected for a heavily doped organic semiconductor. Each current-voltage curve was treated with linear fit (data portion with current over 500 nA which saturated the C-AFM amplifier for voltages between -0.2 V and 0.2 V was discarded) to extract electrical resistance. Figure~\ref{fig:CAFM_map} (e) demonstrates the histogram of logarithm of resistance value (log(R)) fitted with log-normal distribution model, which gives us log-mean value of resistance ($log$-$\mu$) = $5.50$ and log-standard deviation ($log$-$\sigma$) = $0.02$ for this sample area, corresponding to electrical resistance of \SI[separate-uncertainty=true,multi-part-units=single]{316\pm15 }{\kilo\ohm}. This accounts for electrical conductivity $\sigma$ = \SI[separate-uncertainty=true,multi-part-units=single]{51.8\pm2.5}{\siemens\per\centi\meter}. This result is in contrast with values reported for in-plane electrical measurements, where $\sigma$ of PEDOT:OTf films is over 2000 \si{\siemens\per\centi\meter} \cite{Yvenou2020}. In-plane electrical measurements on our materials have provided us with values close to the literature (1500 \si{\siemens\per\centi\meter}, see Section 7 of SI). This large difference in electrical conductivity is attributed to the anisotropy of PEDOT:OTf material\cite{Na2009}. It was previously demonstrated that polymeric chains of \textit{in situ} polymerized PEDOT:OTf have preferential edge-on orientation on the substrate \cite{Gueye2016}. It is well known that semiconducting polymers exhibit better electrical transporting properties in the direction of $\pi-\pi$ stacking \cite{Sirringhaus1999a}, which for PEDOT chains in edge-on orientation lies in the substrate plane. This makes it reasonable to expect electrical conductivity decrease in the cross-plane direction. 
\subsection{Local thermal characterization of pristine PEDOT:OTf films}

Thermal characterization of PEDOT:OTf films was performed with SThM. Figures~\ref{fig:SThM_props} (a) and (b) show topography (in contact mode) and temperature image of a 5$\times$5 \si{\micro\meter} area of a PEDOT:OTf sample with partially exposed gold substrate. For this sample, the film was scratched after deposition with a sharp object, thus creating an area with variable thickness: pristine film with thickness of about 40 to 45 nm, thinner layer of about 8 to 10 nm thick and gold substrate with polymer residue of several nanometers thick (Figure S6 of the SI).



\begin{figure}[h!]
	\centering
	\includegraphics[width=0.99\linewidth]{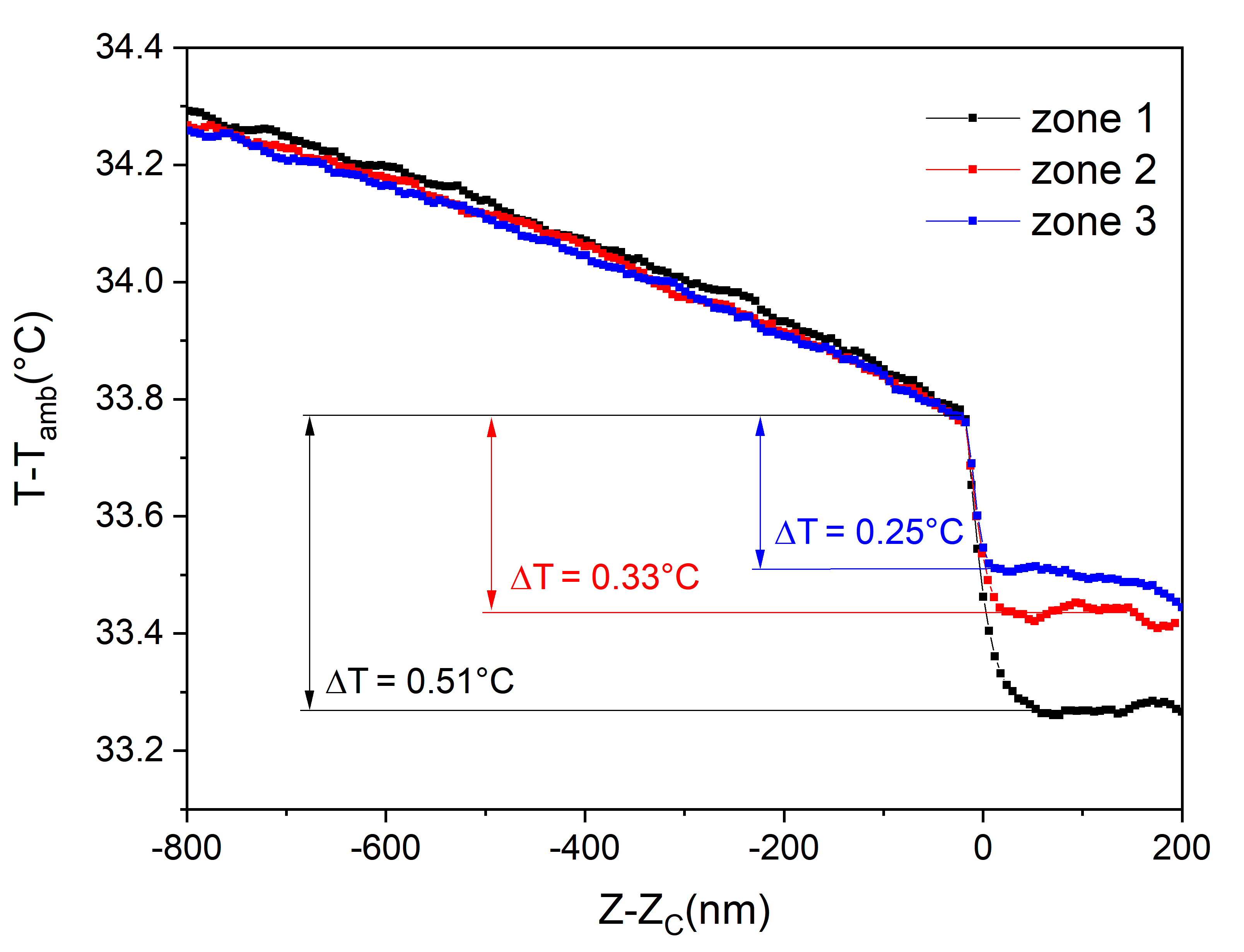}
	\caption[]{Example of NP SThM temperature measurement data (tip temperature as a function of its displacement during the extend ramp): Z$_C$ is scanner coordinate at contact. Wheatstone bridge voltage is 1.1V.}
	\label{fig:pedot_K_layers_dT}
\end{figure}

\begin{figure}[h!]
	\centering
	\includegraphics[width=0.99\linewidth]{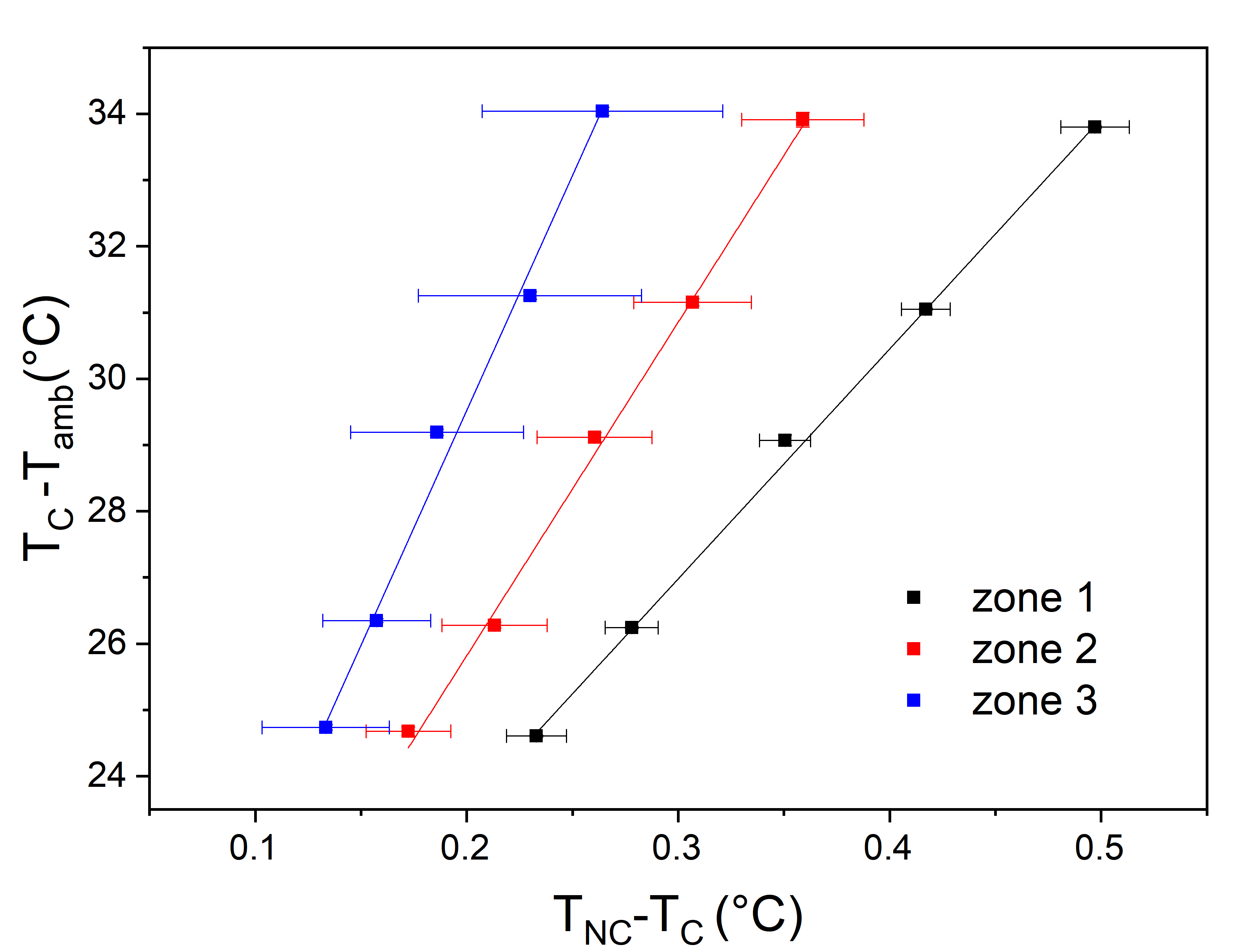}
	\caption[]{Plot of $T_C-T_{amb}$ as a function of $T_{NC}-T_C$ for 3 zones highlighted in the Figure~\ref{fig:SThM_props} (a) and (b), Wheatstone bridge bias was varied between 0.7V and 1.1V in steps of 0.1V. Corresponding thermal conductivity values are presented in the Figure~\ref{fig:SThM_props} (c) and calculated from eq. \ref{eq:np_sthm} using calibration parameters: $\alpha$=38.51 Wm-1K-1 and $\beta$=23.92 K/K (see Figure S1 in SI).}
	\label{fig:pedot_K_layers}
\end{figure}

As we can see from the temperature image (Figures~\ref{fig:SThM_props} (b)), thinner regions of PEDOT:OTf film demonstrate temperature difference. 
For films with thickness of 10 nm and below the contribution of the highly conductive substrate ($\kappa_{Au}$ = 318 \si{\W\per\m\per\K}) results in noticeable temperature contrast (see Figure S7 in SI for temperature profile). We have performed NP SThM measurements on three areas corresponding to the film of varying thickness (square marks on Figures~\ref{fig:SThM_props} (a) and (b)). Measurements were performed in a 4$\times$4 grid with 100 nm pitch per zone, with Wheatstone bridge voltage varied between 0.7V to 1.1V in 0.1V increment. Figure~\ref{fig:pedot_K_layers_dT} demonstrates temperature-distance curves for a single point in each of three zones at 1.1V. The temperature drop upon contact ($\Delta$$T$) is proportional to the thermal conductance. 
For every Wheatstone bridge voltage value per zone, we have extracted values of $T_{NC}-T_C$ and $T_C$ from the extend part of the curve and averaged them over 16 points for the current measurement zone in order to reduce thermal contact area variation. Figure~\ref{fig:pedot_K_layers} demonstrates the plot of $T_C-T_{amb}$ as a function of $T_{NC}-T_C$, which allows us to calculate the apparent thermal conductivity for these three zones (See Figure S1 in SI for calibration data). It is of interest to note the increase of the horizontal error bar ($T_{NC}-T_C$ variation) for thicker sample area which is directly related to contact area variation due to increased surface roughness.

Calculated thermal conductivity as a function of polymer thickness is presented in the Figure~\ref{fig:SThM_props} (c). For the pristine film (thickness about 40 nm), we correct the as-measured value of 0.83 \si{\W\per\m\per\K} by utilizing the model developed by Dryden \cite{Dryden1983} that provides an estimation of the constriction resistance of thin layers. For "thick" ($\frac{t}{r}>2$) films (here 40 nm) we use the following equation (\ref{eq:dryden1}): 
\begin{equation}
	\frac{1}{\kappa^*}=\frac{1}{\kappa_f}-\frac{2}{\pi \kappa_f} \frac{r}{t} \ln\left(\frac{2}{1+\kappa_f/\kappa_{Au}}\right), \label{eq:dryden1}
\end{equation}

where $\kappa^*$ is the measured thermal conductivity, $\kappa_f$ and $\kappa_{Au}$ are thermal conductivity of thin film and gold substrate, respectively,  $r$ is the radius of thermal contact between the SThM probe tip (see Section 10 of SI for details on $r$ estimation) and the sample and $t$ is the film thickness. By taking the film thickness of $40\pm5$ nm, we obtain $\kappa_f$ = $0.64\pm0.03$ \si{\W\per\m\per\K}. Another equation (\ref{eq:dryden2}) was used to correct thermal conductivity values for thinner (here below 10 nm) areas of the film ($\frac{t}{r}<2$): 

\begin{equation}
	\frac{1}{\kappa^*}=\frac{1}{\kappa_{Au}}+\frac{4}{\pi \kappa_f} \frac{t}{r} \left(1-\left(\frac{\kappa_f}{\kappa_{Au}}\right)^2\right). \label{eq:dryden2}
\end{equation}

For measured thermal conductivity in the zones 1 and 2 (3 and 8 nm polymer thickness with 2 nm uncertainty) of 3.57 and 1.46 \si{\W\per\m\per\K}, we get $\kappa_f$ = $0.69\pm0.4$ and $0.74\pm0.22$ \si{\W\per\m\per\K}, respectively (Figure~\ref{fig:SThM_props}, c). 

In "thick" film regime ($\frac{t}{r}>2$), high variation of thickness results in much smaller error for the true value of thermal conductivity than for thinner films whose behavior is dominated by the highly thermally conductive substrate. 

\subsection{Doping degree modulation of PEDOT:OTf}

As-deposited, PEDOT:OTf films may contain up to 28\% of oxidized PEDOT chains \cite{Massonnet2015a}. This high degree of oxidation or "doping" is responsible for the high carrier density, and, consequently, high electrical conductivity. It was demonstrated by Bubnova \textit{et al} \cite{Bubnova2012} that reduction of PEDOT allows to effectively tune the carrier density in conducting polymer films which results in decrease of electrical conductivity and increase of Seebeck coefficient. This motivated us to utilize chemical reduction and oxidation in our study of the electronic part of the thermal conductivity ($\kappa_e$).

\begin{figure}[h!]
	\centering
	\includegraphics[width=0.99\linewidth]{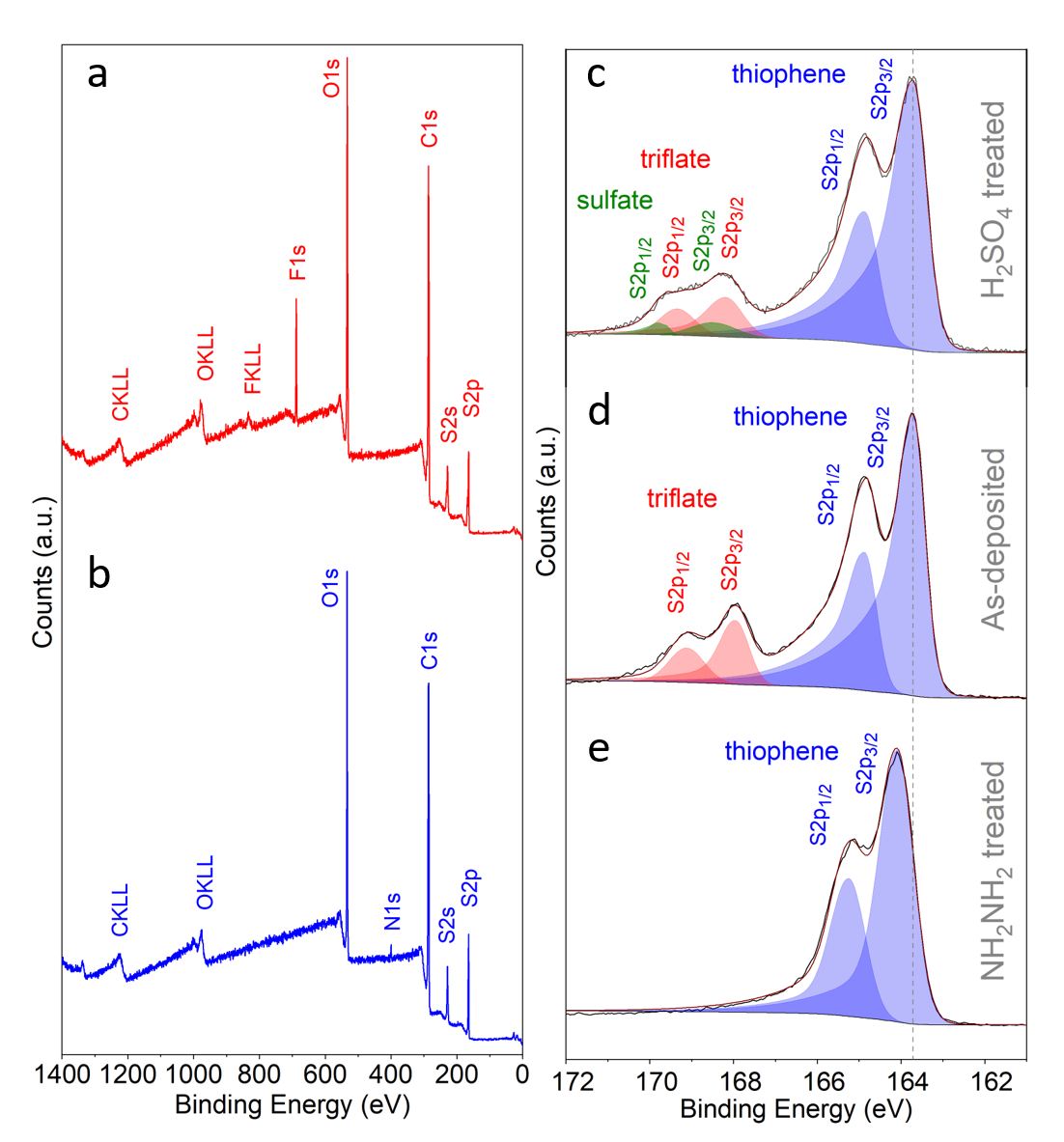}
	\caption[]{XPS spectra of PEDOT:OTf: (a) as-prepared film, (b) after treatment with 10\% wt. hydrazine in ethanol (10 minutes);  S2p region of XPS spectrum: (c) sulfuric acid treated film, (d) pristine film and (d) after reduction. Vertical dashed line corresponds to the energy of S2p$_{3/2}$ of as-deposited PEDOT:OTf. }
	\label{fig:XPS_PEDOT}
\end{figure}

We have modified the procedure described by Massonet \textit{et al} \cite{Massonnet2014} for our materials: pristine PEDOT:OTf films have shown poor wettability with aqueous solutions of reducing agents. We have used solutions of NH$_2$NH$_2$ (NH$_2$NH$_2$$\cdot$H$_2$O in ethanol) and NaBH$_4$ in a mixture of ethanol/water (4:1 v/v) as reducing agents. To reduce PEDOT:OTf, the substrate with deposited polymer layer was immersed in the solution and the exposure time was varied. NaBH$_4$ was used as reducing agent for short reduction times, however it was found unsuitable for prolonged immersion times
: PEDOT layers degrade and delaminate from substrate presumably due to hydrogen gas formation in the bulk of the film. Immersion in hydrazine solutions did not result in mechanical degradation of PEDOT films. The over-doping with H$_2$SO$_4$ was performed by immersing the sample in 1M aqueous solution for 1h. These treatments produced samples with varying doping degree. 

X-ray photoelectron spectroscopy (XPS) was performed to characterize the doping degree control in PEDOT films. The survey spectra for pristine PEDOT:OTf  (Figure~\ref{fig:XPS_PEDOT}, a) shows presence of fluorine (F1s at 688.5 eV) which demonstrates the presence of triflate counter-ions in the film \cite{Massonnet2015a}. The spectrum of hydrazine-treated film reveals the disappearance of the fluorine signal which is indicative of the elimination of triflate counter-ion as a result of de-doping (Figure~\ref{fig:XPS_PEDOT}, b). One can notice a weak signal of nitrogen (N1s at 400.5 eV) in XPS spectrum which may be related to some byproducts of hydrazine oxidation. Data in the S2p region (Figure~\ref{fig:XPS_PEDOT}, c and d) allows to obtain the oxidation degree of pristine PEDOT:OTf by calculating the ratio of thiophene (163-167 eV) and triflate (167-171 eV) signals. The calculated oxidation degree for as-deposited PEDOT:OTf is 19.2\%. Over-doped with sulfuric acid sample demonstrated oxidation degree of 23.7\%. 
Shift of the S2p peak energy to the higher energy in hydrazine-treated PEDOT and the absence of triflate signature corroborates the reduction of thiophene units. It is also of interest to compare the shape of S2p peak of PEDOT: the signals of as-deposited and acid-treated samples demonstrate stronger asymmetry (longer "tail") when compared to the signal of reduced PEDOT. This asymmetry is related to the oxidation degree of PEDOT \cite{Wang2020}: Fabiano \textit{et al.} observed a similar decrease of asymmetry upon chemical reduction of PEDOT:PSS with polyethylenimine vapor \cite{Fabiano2014}.   

\begin{figure}[h!]
	\centering
	\includegraphics[width=0.99\linewidth]{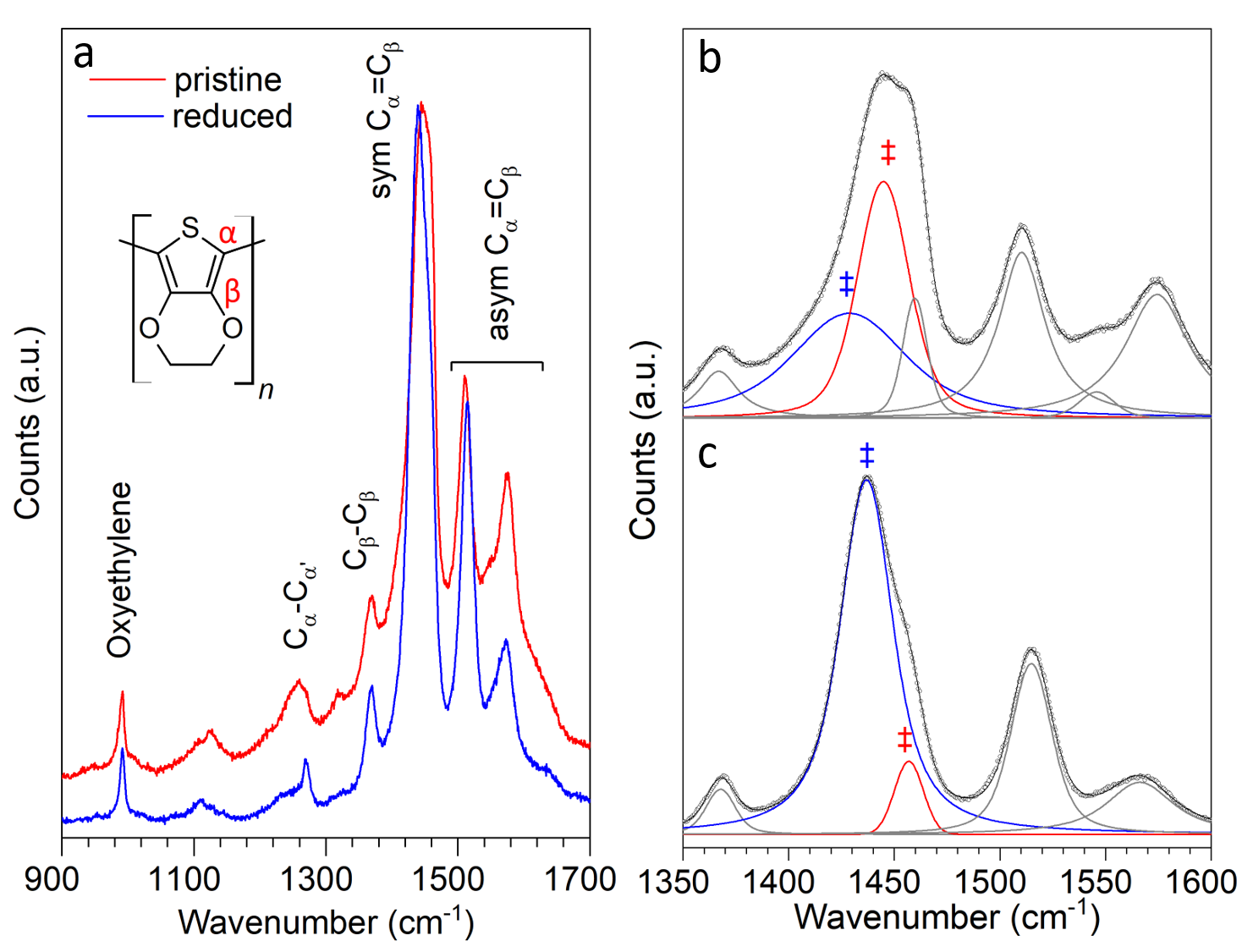}
	\caption[]{ (a) Raman spectra of as-prepared PEDOT:OTf and after treatment with 10\% wt. hydrazine in ethanol (10 minutes). Deconvolution of the spectra  for the pristine (b) and reduced (c) samples; red peak corresponds to the oxidized form, blue - to the neutral form.  }
	\label{fig:Raman_fit_PEDOT}
\end{figure}

Chemical reduction of PEDOT chains results in transition of oxidized thiophene units from quinoid to their aromatic form. Comparison of Raman spectra for pristine and reduced samples is presented in the Figure~\ref{fig:Raman_fit_PEDOT}. It is of interest to note that the spectra recorded for as-deposited samples contain signatures of both neutral and oxidized PEDOT chains, which is demonstrated by the deconvolution of the area from 1350 to \SI{1600}{\per\cm}: vibrational mode related to symmetric C$_{\alpha}$=C$_{\beta}$ stretching (inset of Figure~\ref{fig:Raman_fit_PEDOT}, a) at \SI{1440}{\per\cm} is in fact split in 2 bands (marked with $\ddagger$ in the Figure~\ref{fig:Raman_fit_PEDOT} b and c). As demonstrated by Chiu \textit{et al.} \cite{Chiu2006}, the frequency of these modes is affected by the oxidation degree of PEDOT: lower frequency band represents the contribution of neutral thiophene units (blue) and higher frequency band is related to oxidized form (red). The intensity of the latter appears to be noticeably suppressed in the Raman spectrum of hydrazine-treated PEDOT:OTf, which confirms conversion to neutral form of the majority of PEDOT chains.

\begin{figure}[h!]
	\centering
	\includegraphics[width=0.99\linewidth]{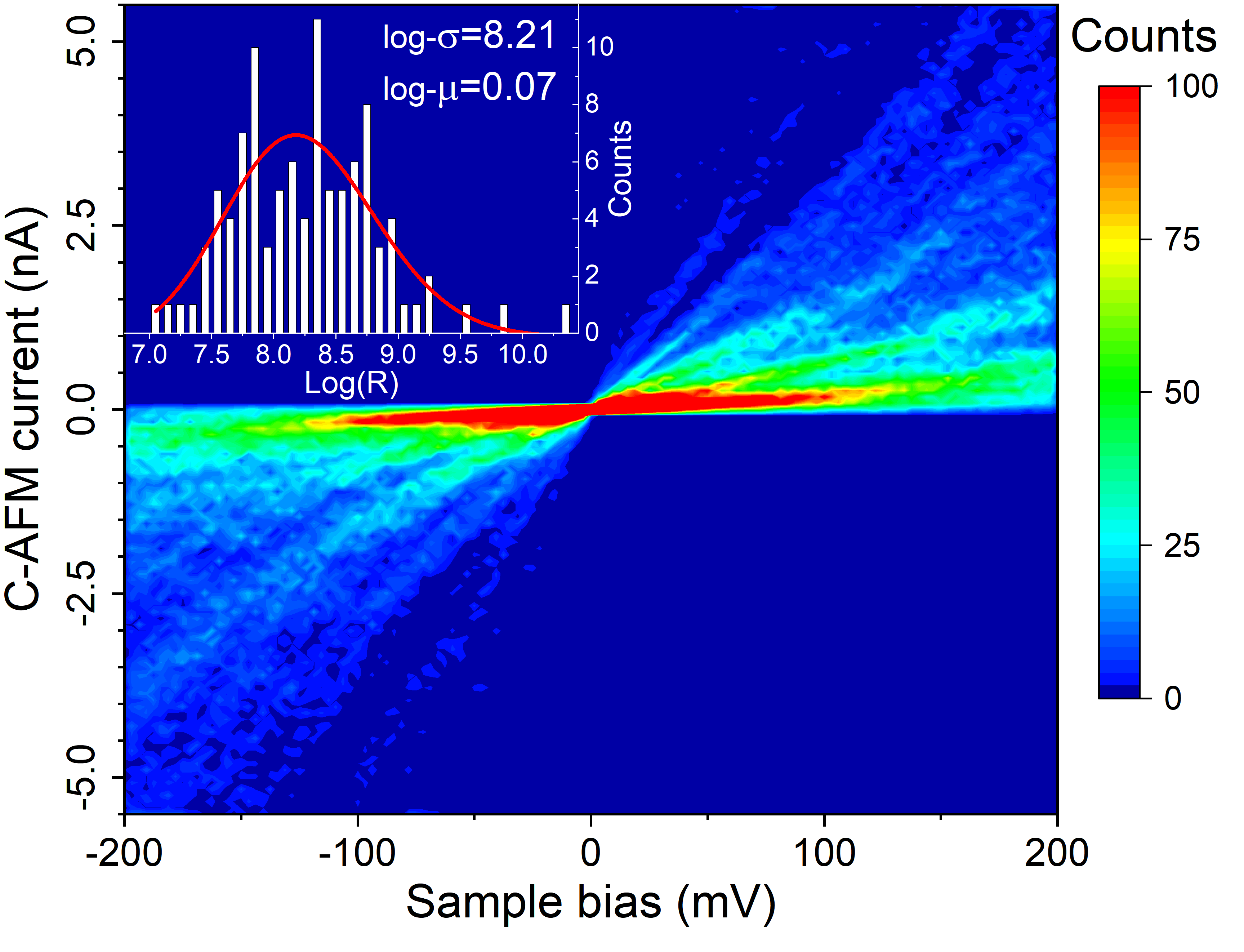}
	\caption[]{2D histogram of I(V) C-AFM measurements of reduced PEDOT (10 minutes in 10\% wt. NH$_2$NH$_2$ solution, 20 nN constant force), 100 measurements in a 10$\times$10 grid with 100 nm pitch. Inset: histogram of Log(Resistance) values extracted from the same data with log-normal distribution fit ($log$-$\mu$=8.21, $log$-$\sigma$=0.07). 
	}
	\label{fig:2d_histo_comp}
\end{figure}    

\begin{figure}[h!]
	\centering
	\includegraphics[width=0.99\linewidth]{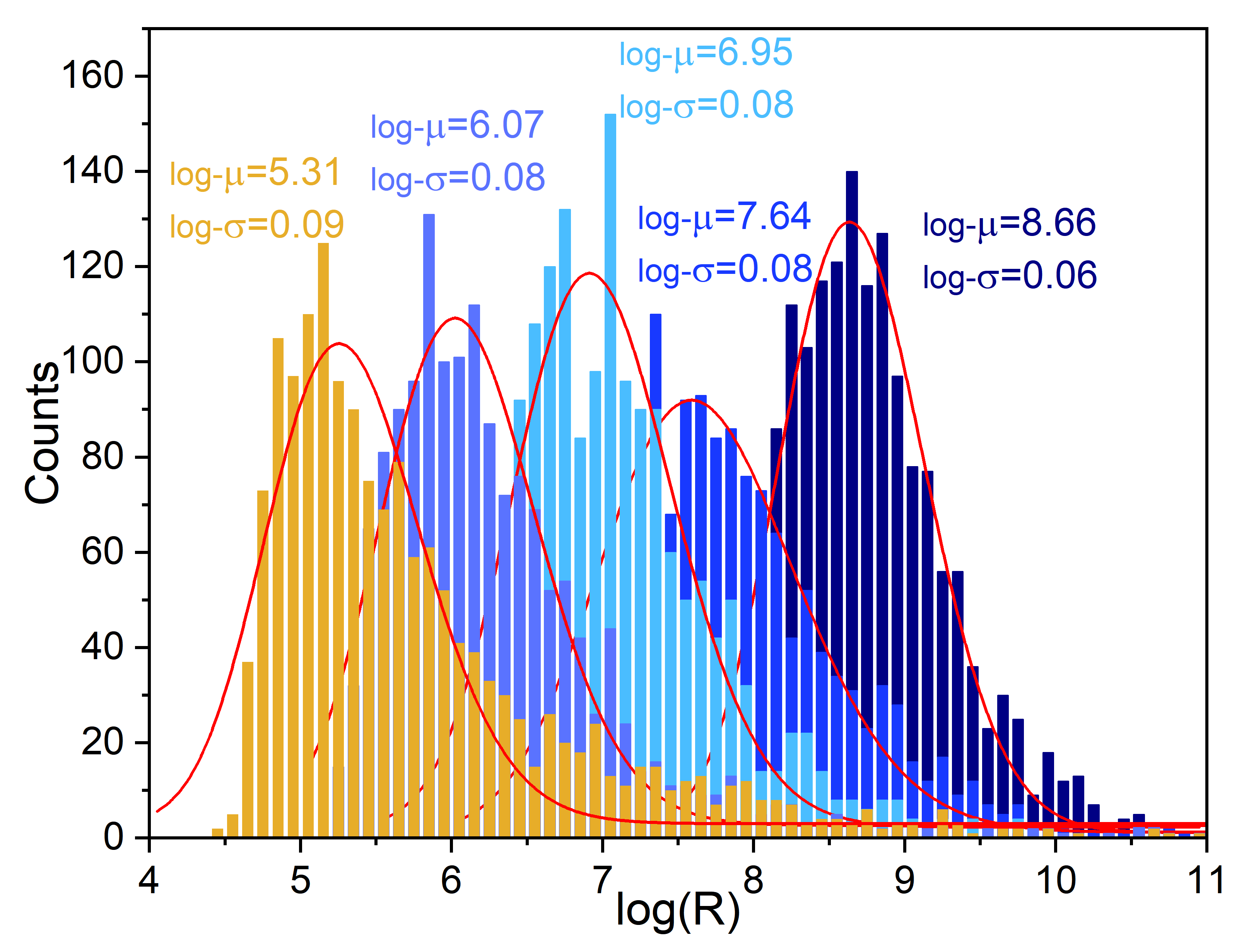}
	\caption[]{Histograms and log-normal fits of log(Resistance) extracted from C-AFM measurements. 1600 I(V) plots are used for each histogram. From left to right: PEDOT:OTf prepared with 7.75\% wt. DMF co-solvent after H$_2$SO$_4$ treatment: no co-solvent; 5 minutes in 10\% wt. NaBH$_4$ solution; 10 minutes in the same solution and 30 minutes in 10\% wt. NH$_2$NH$_2$ solution.}
	\label{fig:logR_red}
\end{figure}



Reducing treatment has pronounced effect on electrical conductivity: decrease of the doping degree of PEDOT:OTf lowers the concentration of charge carriers and electrical conductivity. A sample with 10 minutes exposure to 10\% wt. hydrazine solution exhibits strong decrease of current amplitude 
(Figure~\ref{fig:2d_histo_comp}). 
Inset of Figure~\ref{fig:2d_histo_comp} shows the log-normal distribution of electrical resistance values (fitted values correspond to resistance R = \SI[separate-uncertainty=true,multi-part-units=single]{162\pm26 }{\mega\ohm}
) which corresponds to \SI[separate-uncertainty=true,multi-part-units=single]{0.106\pm0.015 }{\siemens\per\centi\meter} (about three orders of magnitude lower than for pristine sample). 

By utilizing different immersion times and concentration of reducing agents as well as over-doping in sulfuric acid as described by Massonnet \textit{et al.} \cite{Massonnet2015a}, we were able to vary the electrical conductivity of PEDOT samples in the range between 0.04 to 70  \si{\siemens\per\centi\meter}. Figure~\ref{fig:logR_red} shows comparison of log-normal distributions of electrical resistance for samples obtained after various post-deposition treatments, see the legend for details. Reduced PEDOT demonstrated quasi-ohmic behavior which allowed us to extract resistance values from 1600 I-Vs per sample (400 I–V recorded in a 20$\times$20 grid with pitch of 25 nm on randomly selected 4 zones on the polymer film, see Figure S8 for 2D histograms of current values distribution).




\subsection{Correlation of thermal conductivity with electronic transport}

\begin{figure}[h!]
	\centering
	\includegraphics[width=0.99\linewidth]{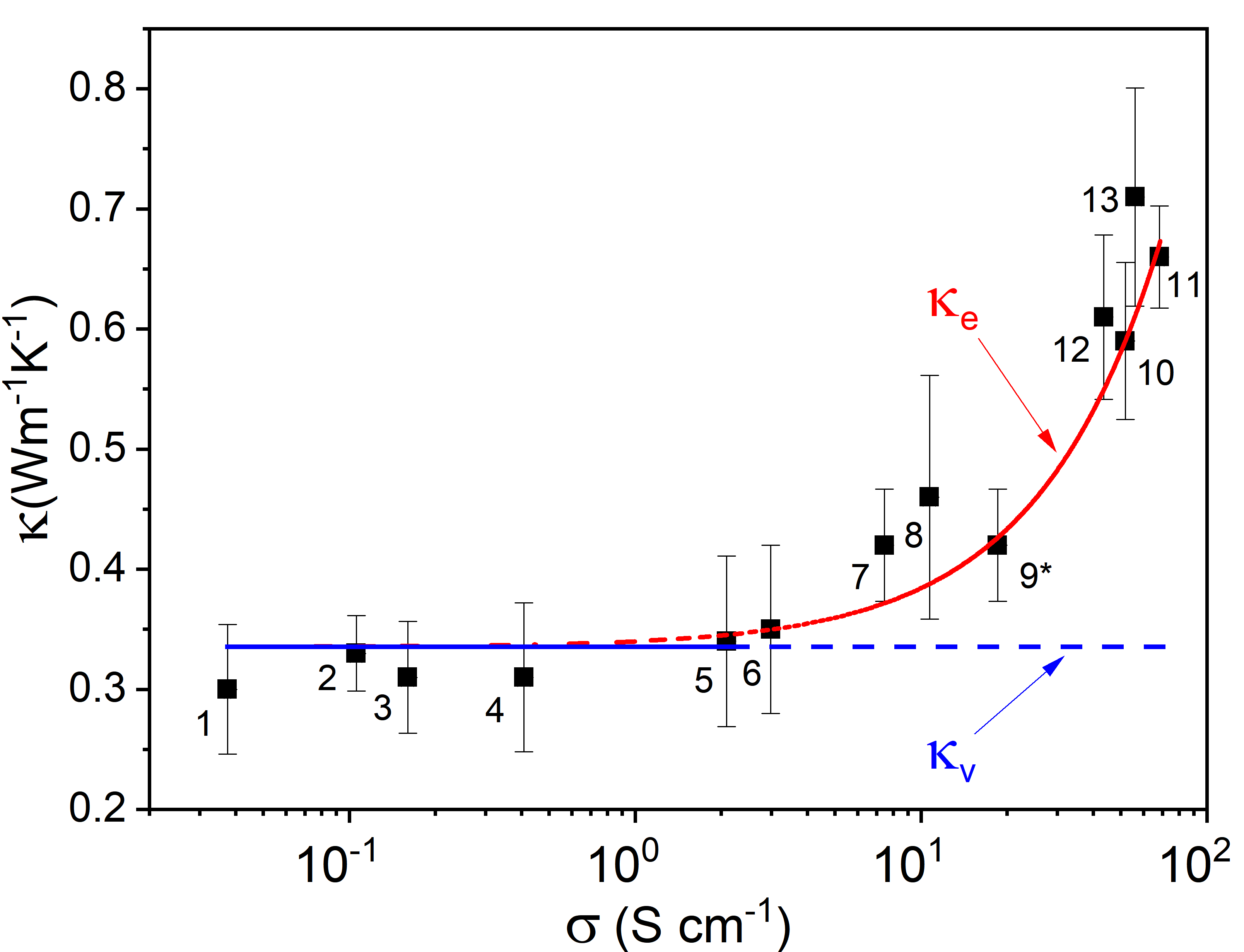}
	\caption[]{Plot of thermal conductivity as a function of electrical conductivity of PEDOT:OTf. Sample 1 was treated with 10\% wt. NH$_2$NH$_2$ for 30 minutes, sample 2 - 10 minutes, sample 3 - 5 minutes. Samples 4 to 8 were treated with NaBH$_4$ solution: for samples 4 to 6, exposure to 10\% wt. solution was 10, 5 and 2 minutes, samples 7 and 8 were treated with 2\% wt. concentration for 10 and 5 minutes, respectively. Sample 9 represents PEDOT:OTf with no co-solvent in the precursor mixture, samples 10 and 11 were prepared with 7.75\% wt. DMF (before and after H$_2$SO$_4$ treatment) and samples 12 and 13 included 7\% wt. NMP in their formulation (before and after H$_2$SO$_4$ treatment), respectively. }
	\label{fig:K_vs_logS}
\end{figure}

The cross-plane thermal conductivity of the samples described in the previous subsection was further measured by the NP-SThM technique. Measurements were performed on 5 randomly selected spots of the sample. As-measured thermal conductivity values were corrected in relation to substrate (gold) contribution according to the procedure presented in the section~\ref{section:Nanoscale_props} (see Section 12 of SI for NP SThM data).

It was found that thermal conductivity increases with the electrical conductivity of the sample (Figure~\ref{fig:K_vs_logS}). For samples with low electrical conductivity (below 1 \si{\siemens\per\centi\meter}) the thermal conductivity demonstrates negligible influence of the electronic transport, which allows us to extract the vibrational thermal conductivity $\kappa_v$ = 0.34±0.04 \si{\W\per\m\per\K}.

\begin{figure}[h!]
	\centering
	\includegraphics[width=0.99\linewidth]{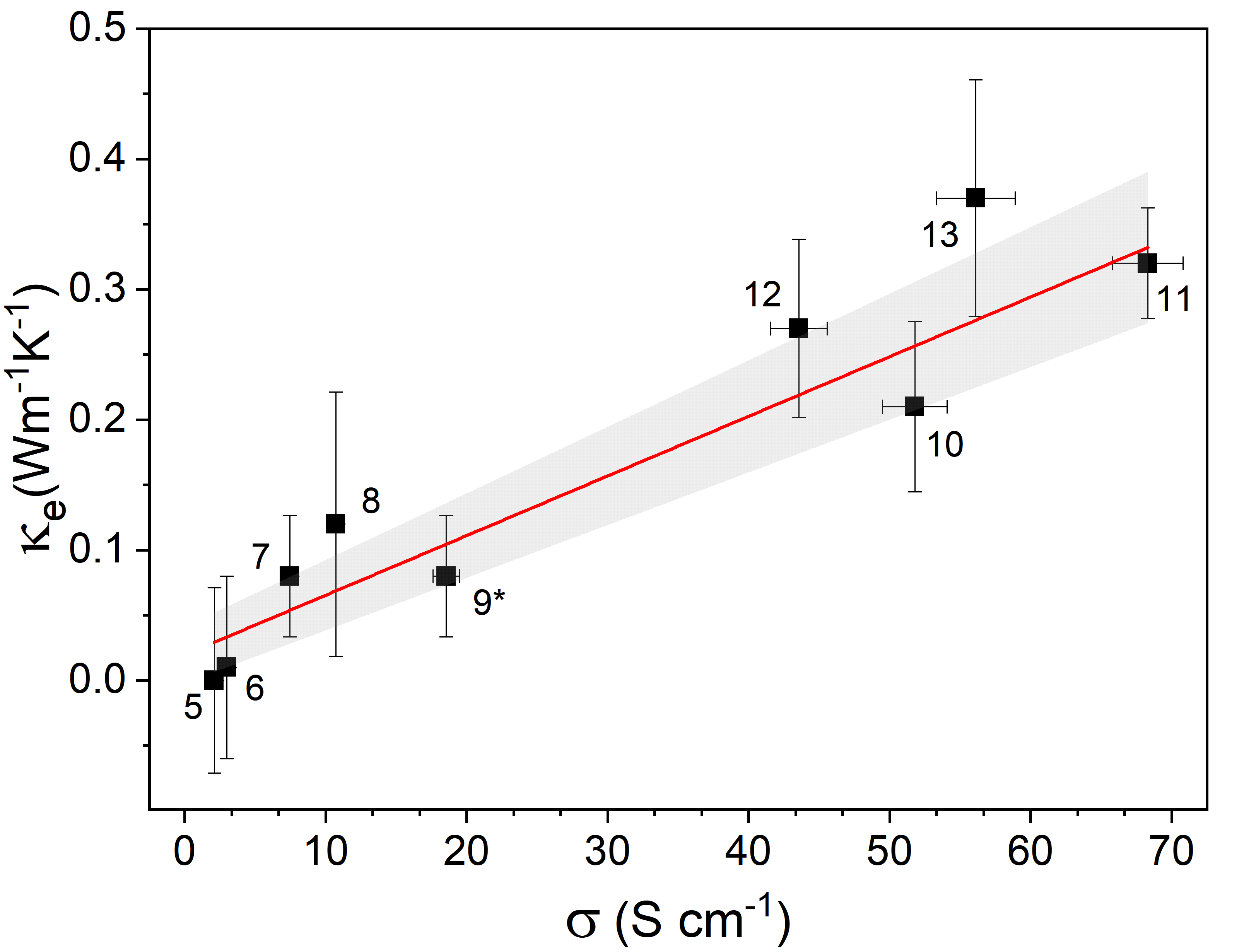}
	\caption[]{Plot of electrical contribution to thermal conductivity of PEDOT:OTf. Sample numbers correspond to the legend of the Figure~\ref{fig:K_vs_logS}. Linear fit is in red, gray area corresponds to the fit error.}
	\label{fig:K_electronic}
\end{figure}


We have plotted the electronic contribution to thermal conductivity ($\kappa _e$=$\kappa$-$\kappa_v$) corresponding to the samples with electrical conductivity over 1 \si{\siemens\per\centi\meter}) in Figure~\ref{fig:K_electronic}. This data has allowed us to asses the applicability of classic Widemann-Franz law (Eq. \ref{eq:WF_law}) to the cross-plane transport in PEDOT:OTf system: 
\begin{equation}
	\kappa _e = L \sigma T , \label{eq:WF_law}
\end{equation}
where $L$ is the Lorenz number. From our data, we extract $L$ = \SI[separate-uncertainty=true,multi-part-units=single]{1.54\pm0.18e-7}{\watt\ohm\per\kelvin\squared}. This value is about 6 times larger than classic Sommerfeld value ( \SI{2.44e-8}{\watt\ohm\per\kelvin\squared}).
We have also made an attempt to apply recently developed molecular WF law (Eq. \ref{eq:MWF_law}) \cite{Craven2020}, which instead of $L$ relates the $\kappa _e$($\sigma$) dependence to the reorganizational energy of the molecule $\lambda$ (in \si{\eV}):
\begin{equation}
	\kappa _e = k_B \lambda \sigma, \label{eq:MWF_law}
\end{equation}
where $k_B$ is the Boltzmann constant (in \si{\electronvolt\per\K}). We obtain $\lambda$ = \SI[separate-uncertainty=true,multi-part-units=single]{0.53\pm0.06}{\eV} from our data (Figure~\ref{fig:K_electronic}).

\section{Discussion}

PEDOT-based materials are among the most studied conductive polymers due to the vast variety of works dedicated to the investigation of their electrical and thermal transport. It was demonstrated by Br\'edas \textit{et al.} \cite{Kim2008} and Xi \textit{et al} \cite{Shi2015} from the band structure of PEDOT stabilized with toluenesulfonyl (Tos) ions that the conduction in this material is essentially 2D (inside of the PEDOT lamellae: along the polymer chain and in $\pi$-$\pi$ stacking direction). This information corroborates the observation of lower cross-plane electrical conductivity of PEDOT:OTf material, as due to the chain orientation on the substrate (edge-on) our measurements are carried out in inter-lamellar direction.



We have investigated charge and heat transport in PEDOT:OTf system as a function of oxidation degree. These approaches have been successfully applied in the past by Bubnova \textit{et al} \cite{Bubnova2012} in order to tune the carrier density of PEDOT and find a suitable trade-off between the decrease of electrical conductivity and the increase of Seebeck coefficient in the numerator of the equation (\ref{eq:ZT}). It was primarily assumed that decrease of electrical conductivity is governed by the decrease of charge carrier density with the mobility not being significantly affected by the aforementioned treatment \cite{Lee2014c}, 
however, it is important to take into consideration dopant-induced changes of crystalline structure \cite{Yee2019} to fully understand the behavior of $\kappa_v$. 
Small ion dopants in PEDOT materials are intercalated in between the lamellar structure of PEDOT chains \cite{Shi2015}, \cite{Kim2008}. In fact, insertion of these small ions significantly modifies the crystalline structure of the material: it results in shrinkage of the unit cell of PEDOT  crystal (promoting $\pi$-$\pi$ interactions) and an increase of inter-lamellar distance \cite{Shi2015}. 
As we have mentioned above, our measurements correspond to inter-lamellar direction of transport. Thus, it is reasonable to expect the phonon thermal conductivity to be affected by the presence of ionized inclusions of dopant \cite{Wang2012b}: it is well known that such impurities increase phonon scattering \cite{Zapata-Arteaga2020}. This implies that \textit{true} $\kappa_v$ may be not constant with the electrical conductivity (doping degree) and caution should be exerted upon subtraction of $\kappa_v$ from the $\kappa(\sigma)$ linear dependence.

The value of cross-plane lattice thermal conductivity shown in the Figure~\ref{fig:K_electronic} is in agreement with the observations of Liu \textit{et al} \cite{Liu2015f} (0.3 \si{\W\per\m\per\K}), where he demonstrated significant anisotropy (factor of 2 between in-plane and cross-plane directions) which is analogous to electrical transport in PEDOT. In his study, he demonstrates that cross-plane $\kappa$ is independent of in-plane $\sigma$ (cross plane $\sigma$ was not measured), but in-plane $\kappa$ follows the classic WF law. 
One may argue that since the oxidation degree was not monitored, it is unclear if the encountered $\sigma$ increase is due to the variation of ionized dopant content and thus allows the authors to suggest from their elastic constant measurements that the lattice parameters are not changed (as well as $\kappa_v$). It was suggested by Lee \textit{et al} \cite{Kim2014} that in PEDOT:PSS material, PEDOT chains prefer lamellar stacking pattern with insulating PSS located in between the lamellae. This may help to explain the behavior of cross-plane $\kappa$ from the measurement of Liu et al \cite{Liu2015f}, since it is common for this material to have large fraction of insulating PSS polymer with $\kappa$ of 0.38 \si{\W\per\m\per\K} \cite{Kim2016}.

Numerous studies of thermoelectric transport in organic semiconductors have investigated $\kappa_e$ in the framework of classical Wiedemann-Franz law, which was originally developed for metals. Though it provides reliable interpretation of $\kappa_e(\sigma)$ trend in metals and some semiconductors, it was found that for systems with pronounced charge carrier localization such as disordered organic semiconductors, the $\kappa(\sigma)$ significantly deviates from expected behavior. It was demonstrated that for semiconducting polymers, the  deviation from classic value of Lorentz number $L$ in WF law can surpass the initial value by 10 times: polyaniline demonstrates $L$ higher by an order of magnitude than Sommerfeld value \cite{Yoon1991}. Similar observations for in-plane $\kappa_e$ of PEDOT:Tos prepared by vacuum vapor phase polymerization were reported by Weathers \textit{et al.}: value of $L$ was 2.3 times higher than expected for classic WF behavior \cite{Weathers2015}. Xu \textit{et al.} have compared Lorenz number values encountered for conductive polymers which allowed them to propose that it depends strongly on the doping degree of the polymer and this dependence is the direct consequence of the quasi-particle (polarons and solitons) nature of transport in conductive polymers \cite{Xu2020}. 

The uncertainty of applicability of classic WF law to organic semiconductors has led to the development of a novel theory connecting thermal and electrical transport in molecular systems by Craven and Nitzan \cite{Craven2020}. Hopping between charge sites is governed by the molecular reorganization energy $\lambda$, which essentially replaces the $LT$ product from the classic WF law (\ref{eq:WF_law}). 
It is difficult to asses the validity of our $\lambda$ estimation (\SI[separate-uncertainty=true,multi-part-units=single]{0.53\pm0.06}{\eV}) for PEDOT:OTf, given the usual $\lambda$ values for model polyacene semiconductors below 0.2 eV \cite{Duhm2012}, \cite{Kera2015}. Oliveira \textit{et al} have estimated $\lambda$ for poly-alkyl-thiophene based polymers to be in the order of 0.2-0.9 eV \cite{Oliveira2016}. Theoretical calculations of $\lambda$ very often neglect the contribution of external $\lambda$ (i.e. the contribution of the surrounding environment, which is reasonable for undoped materials), however, external effects could not be overlooked for a material with significant ionized dopant content such as PEDOT:OTf. Total reorganization energy is strongly dependent on the molecular environment \cite{Hernangomez-Perez2020} and the computational costs of modeling of such systems are often prohibitive. Additional work is required to provide a more appropriate theoretical estimate of $\lambda$, which is out of scope of the present publication.
\section{Conclusion}

In this work, we present the study of charge and heat transport in the cross-plane direction of PEDOT:OTf material at the nanoscale. We have explored the link between the morphology and electrical properties of our material: the polymer presents highly conductive nanodomains which correspond to the grains of highly ordered PEDOT chains. Triflate-doped PEDOT presents ohmic behavior which has allowed us to obtain cross-plane electrical conductivity of the as-deposited polymer in the order of \SI{50}{\siemens\per\centi\meter}. 
We have further investigated thermal conductivity in PEDOT:OTf thin films: the behavior of cross-plane thermal conductivity as a function of film thickness follows a simple analytical model, which allowed us to effectively decouple this property from thermally conductive substrate; for films of 40 nm and thicker the variation of thickness of $\pm$5 nm results in experimental uncertainty no higher than 10$\%$.

The correlation between cross-plane thermal and electrical transport of PEDOT:OTf is also reported in this work. Chemical reduction and acid treatment allowed us to follow the evolution of electronic contribution to thermal transport on a range of more than three decades of electrical conductivity. We have found that in weakly conducting PEDOT, the electronic contribution $\kappa_e$ is negligible and the thermal conductivity is characterized by its vibrational term $\kappa_v$ = 0.34±0.04 \si{\W\per\m\per\K}. In highly electrically conductive PEDOT samples, the electronic contribution $\kappa_e$ reaches the same order of magnitude as $\kappa_v$.

The Lorenz number value for cross-plane transport in PEDOT:OTF which we have estimated from our data is about 6 times higher than classic Sommerfeld value, which raises questions on applicability of WF law to this class of materials despite almost metallic character of in-plane transport in PEDOT:OTf. 
We deduce reorganization energy of $\lambda$ =  \SI[separate-uncertainty=true,multi-part-units=single]{0.53\pm0.06}{\eV} from the newly developed molecular WF law, a value slightly higher than expected that calls for more theoretical studies and experimental confirmations.

\section{Conflicts of interest}

The authors declare no conflict of interest.

\section{Acknowledgment}

We thank the financial support from French National Research Agency (ANR): project HARVESTERS, ANR-16-CE05-0029. The IEMN facilities are partly supported by Renatech. We thank D. Deresmes for his valuable help with the SThM
instrument.
	
	\renewcommand\refname{References}
	
	\bibliography{library} 
	\bibliographystyle{rsc} 
	
\end{document}